\documentstyle[12pt,amsfonts]{article}

\def\NPB#1#2#3{Nucl. Phys. B {\bf#1} (19#2) #3}

\def\s2{\frac{1}{\sqrt2}}

\def\oh{\frac{1}{2}}
\def\beq{\begin{equation}}
\def\eeq{\end{equation}}
\def\beqa{\begin{eqnarray}}
\def\eeqa{\end{eqnarray}}
\def\Z{{\Bbb Z}}
\def\tr{{\rm tr \,}}
\def\Tr{{\rm Tr \,}}
\def\deq#1{\mbox{$d$=#1}}

\def\r#1{\mbox{{\bf #1}}}

\begin{document}

\pagestyle{empty}
\begin{flushright}
SB/F/02-298 \\
{\tt hep-th/mmddyy}
\end{flushright}
\vspace*{2cm}

\vspace{0.3cm}
\begin{center}
\LARGE{\bf Non-Supersymmetric Orbifolds }\\[1cm]
{\large Anamar\'{\i}a Font}${}^*$\footnote{afont@fisica.ciens.ucv.ve} 
{\large and Alexis Hern\'andez}${}^\dag$
\footnote{ahernand@fis.usb.ve} \\[2mm]
${}^*${\normalsize  Dept. de F\'{\i}sica, Fac. de Ciencias,
Universidad Central de Venezuela,}\\[-5mm]
{\normalsize A.P. 20513, Caracas 1020-A, Venezuela. }\\[1mm]
${}^\dag${\normalsize Dept. de F\'{\i}sica, 
Universidad Sim\'on Bol\'{\i}var,}\\[-5mm]
{\normalsize A.P. 89000, Caracas 1080-A, Venezuela.}\\[1cm]
\small{\bf Abstract} \\[3mm]
\end{center}

\begin{center}
\begin{minipage}[h]{15.5cm}
\small{
We study compact non-supersymmetric
$\Z_N$ orbifolds in various dimensions. We compute the spectrum 
of several tachyonic type II and heterotic examples and
partially classify tachyon-free heterotic models. We also discuss
the relation to compactification on K3 and Calabi-Yau manifolds. }
\end{minipage}
\end{center}

\newpage

\setcounter{page}{1} \pagestyle{plain}
\renewcommand{\thefootnote}{\arabic{footnote}}
\setcounter{footnote}{0}

\section{Introduction}
Orbifold compactifications provide a large class of exactly solvable
string vacua in which the number of supersymmetries is reduced \cite{dhvw}.
In this paper we will discuss type II and heterotic orbifolds in which
supersymmetry is completely broken by the twisted boundary conditions.
This generalizes the Harvey-Dixon construction \cite{dh} of the ten 
dimensional non-supersymmetric strings that can be found in various other
ways \cite{allns}. 

To our knowledge, the study of non-supersymmetric compact orbifolds 
has not been carried out systematically until now. Compactifications
of the \deq10 non-supersymmetric theories were considered some time ago 
\cite{allcom}. Non-supersymmetric closed
string models in lower dimensions have
been devised using asymmetric constructions \cite{fourcom}, the 
Green-Schwarz formulation \cite{cdq}, Scherk-Schwarz 
compactifications \cite{scherk}
and compactifications on magnetic backgrounds \cite{russo}.
The question of vanishing of the cosmological constant 
in non-supersymmetric strings has been addressed as well \cite{lambda}.

In the non-supersymmetric orbifolds that we discuss,
supersymmetry is broken at the string scale. Thus, potential 
phenomenological applications in \deq4 will face the hierarchy problem. 
Now, an approach to solve this problem in a similar scenario
has actually been proposed \cite{dienes} and other mechanisms could be 
conceived in the future. At any rate, since experimental evidence 
indicates that supersymmetry is broken, non-supersymmetric strings
should be explored. Moreover, these theories have interesting features
that are worth studying in more depth. Notably, such theories typically
have tachyons and are thus unstable. It is then necessary to investigate
processes of vacuum stabilization. Recently, condensation of closed type II
tachyons has been analyzed in the case of non-compact orbifolds \cite{cond}.
The fate of tachyons in compact orbifolds has however not been
considered in detail. The case of heterotic tachyons has only been
discussed in \deq10 \cite{suyama}. In this paper we will 
present several tachyonic examples in $d=8,6,4$, that could serve 
as starting points to study type II and heterotic
tachyon condensation in compact orbifolds.
Furthermore, tachyons could be of relevance in electroweak 
symmetry breaking \cite{higgst} and in inflation \cite{bmnqrz}.
Yet another interesting issue is that of duality of string theories without
supersymmetry. We have found several non-tachyonic heterotic models
whose strong coupling duals could be described as in the
$SO_{16}\times SO_{16}$ case \cite{blum}.

This paper deals with type II and heterotic non-supersymmetric 
compactifications on $T^{10-d}/\Z_N$ for $d=8,6,4$ and $N \leq 6$.
Our main results consist of concrete examples. In particular, we have
determined all tachyon-free heterotic models for $N \leq 5$. As a check
on the massless spectra we have verified factorization of anomalies.
We have also found relations to compactifications of the \deq10
non-supersymmetric strings on smooth K3 and Calabi-Yau manifolds.

This paper is organized as follows. In section 2 we introduce the basic 
notation and determine the allowed $\Z_N$, $N \leq 6$, twists that
break supersymmetry. In sections 3 and 4 we describe type II and heterotic
orbifolds. In section 5 we present our final comments. For the sake of
self-containedness we include an appendix with a review of tools needed 
to obtain the tachyonic and massless spectrum in compact orbifolds.

\section{Generalities}

We focus on $T^{10-d}/\Z_N$ orbifolds although the analysis can be easily
extended to $\Z_N \times \Z_M$ actions and discrete torsion can also be
included. In our standard notation the $\Z_N$ generator, denoted $\theta$,
acts diagonally on internal complex coordinates as 
$\theta Y_i = e^{2\pi i v_i} Y_i$. Since $\theta^N=1$, the $v_i$
are of the form $k/N$. The $v_i$ are
restricted by the condition that $\theta$ must act crystallographically
on the torus lattice. In particular this requires that the number
of fixed points $\det (1-\theta)$ be an integer. All such twists
have been classified in ref.\cite{ek}. Spacetime supersymmetry imposes
the extra condition
\beq
\pm v_1 \pm v_2 \pm v_3 = 0\ {\rm mod}\  2
\label{susyc}
\eeq
Modular invariance gives the additional constraint
\beq
N \sum_i v_i = 0\ {\rm mod}\  2
\label{modc}
\eeq
Notice that (\ref{modc}) also ensures that $\theta$ is of order
$N$ acting on the world-sheet fermionic degrees of freedom.

All crystallographic twists satisfying (\ref{susyc}) and (\ref{modc})
were given in ref.\cite{dhvw}. If we relax (\ref{susyc}) there will
be no gravitini left in the untwisted sector so that supersymmetry 
is generically broken.
In this case we can still use the results of ref.\cite{ek} to find
the allowed $(v_1,v_2,v_3)$. The results for $N \leq 6$ are shown in 
Table \ref{tab1}. For greater $N$ there are of course other solutions.
Notice that in some $\Z_{2M}$ examples, the generator
$\theta$ leaves the holomorphic two or three form invariant
so that the resulting orbifold is either a singular K3 or a 
singular Calabi-Yau. Hence, we will be able to describe a class 
of non-supersymmetric compactifications
in the orbifold limit of these manifolds.

\begin{table}[htb]
\renewcommand{\arraystretch}{1.25}   
\begin{center}
\begin{tabular}{|c|c||c|c|}
\hline
$\Z_N$ & $(v_1,v_2,v_3)$ & $\Z_N$ & $(v_1,v_2,v_3)$  \\
\hline
$\Z_2$ & (0,0,1) & 
$\Z_6$ & (0,0,$\frac13$) \\
\cline{1-2}
$\Z_3$ & (0,0,$\frac23$) &
{} & (0, $\frac16$, $\frac12$)  \\
\cline{1-2}
$\Z_4$ & (0, 0, $\frac12$)  &
{} & (0, $\frac16$, $\frac56$) \\
{} & (0, $\frac14$, $\frac34$)  &
{} & (0, $\frac13$, $\frac23$)  \\
{} & ($\frac12$, $\frac12$, $\frac12$)& 
{} & ($\frac16$, $\frac16$, $\frac23$)  \\
\cline{1-2}
$\Z_5$ & (0, $\frac15$, $\frac35$) &  
{} & ($\frac13$, $\frac13$, $\frac13$)  \\
{} & {}  & 
{} & ($\frac12$, $\frac12$, $\frac13$) \\
\hline
\end{tabular}  
\end{center}
\caption{Twist vectors for non-supersymmetric $\Z_N$ actions, $N \leq 6$.}
\label{tab1}
\end{table}

It is also necessary to specify the torus lattice that allows a
given action. We mostly consider products of two-dimensional
sub-lattices. More precisely, for order two and order four rotations
we take the $SO_4$ root lattice whereas for order three and order six
rotations we take the $SU_3$ root lattice. The $\Z_5$ action
is realized as the Coxeter rotation on the $SU_5$ root lattice.

Several of the twist vectors in Table \ref{tab1} have been considered
previously. The $\Z_2$ orbifold was originally analyzed in the 
context of non-supersymmetric 10-dimensional strings \cite{dh}. 
The $\Z_5$ has been considered in ref. \cite{ks} in 
relation to type IIB orbifolds of $AdS_5 \times S^5$. 
The $\Z_4$ with eigenvalues $(\oh, \oh, \oh)$ has been
used in \cite{bfl} and \cite{kns} to discuss type 0B orbifolds
and in \cite{bk} to construct non-supersymmetric tachyon-free
orientifolds. The $\Z_4$ with  $(0,0,\oh)$ 
has also been used to build a type IIB orbifold \cite{kns}. 
More recently, some of these generators have been considered
in the case of {\it non-compact} orbifolds \cite{cond}.

\section{Type II}

Symmetric type II orbifolds with the non-supersymmetric twists in
Table~\ref{tab1} will
have tachyons in some twisted sector. Notice that in some of the
$\Z_{2M}$ examples, the $M$-th twisted sector is actually
`untwisted' in the sense that $\theta^M$ acts trivially in
spacetime. In such cases the $\theta^M$ sector always
includes a tachyon that could be referred to as untwisted.
Since in these examples $\theta^M= (-1)^{F_S}$, where $F_S$ is
the spacetime fermion number, there are altogether no 
spacetime fermions in the orbifold spectrum. In particular,
the $\Z_2$ twist $(0,0,1)$ leads to the ten-dimensional type 0 closed strings. 
Below we will mostly describe examples with spacetime fermions.

\subsection{$\boldmath{d=8}$}

In this case the little group $SO_8$ is broken to $SO_6\times SO_2$, 
with $\r8_v = (\r6,0) + (\r1,1) + (\r1,-1)$ and
$\r8_s = (\r4,-\oh) + ({\overline{\r4}}, \oh)$, where the second entry
gives the $U_1 \sim SO_2$ charge. 

As an example let us take the $T^2/\Z_3$ with $v=(0,0,0,\frac23)$. 
Using the results shown in the appendix, we see that
the massless states in the untwisted sector include
the metric, dilaton and antisymmetric tensor. In the type IIB case, 
the remaining tachyonic and massless matter is given by
\beq
\begin{array}{ll}
\theta^0: & \r1 + \r1 + \r{15} + \r4 + \r4 \\[0.2ex]
\theta: & 3(\r1^-) + 3 (\r1 + \r1 + \r{15} + \r4 + \r4 )
\end{array}
\label{spz3b}
\eeq
where states are labelled by their $SO_6$ representations and 
$\r1^-$ denotes a tachyon. The three tachyons (one per each fixed point)
in the $\theta$ sector have $m_R^2= m_L^2=-\frac13$. The massless
states in the $\theta$ sector all have oscillators acting on
the twisted vacuum. To the above we must add the antiparticles.

In type IIA, in the untwisted sector there appears instead
\beq
\r1 + \r6 + \r{10} + \r4 
+ {\overline{\r4}}
\label{dosapr}
\eeq 
and in the twisted sector the same with multiplicity three.

\subsection{$\boldmath{d=6}$}

In this case we can classify the massless states according to the little 
group $SO_4$. We use $SO_4 \sim SU_2 \times SU_2$ so that the vector
$(\underline{\pm 1, 0})$ becomes the $(\oh, \oh)$ representation, the spinor
$(\underline{\oh, -\oh})$ becomes $(\oh,0)$ and $\pm (\oh, \oh)$ is $(0,\oh)$.
In the untwisted sector there always appear massless states in the
$SU_2 \times SU_2$ representations $(1,1) + (0,0) + (1,0) + (0,1)$
that correspond to the metric, dilaton and antisymmetric tensor.
The remaining tachyonic and massless spectrum depends on the particular
$\Z_N$. 

As a first example we take the $T^4/\Z_4$ with $v=(0,0,\frac14,\frac34)$.
In the $\theta$ and $\theta^3$ sector there appear tachyons, 
denoted $(0,0)^-$, with $m_R^2= m_L^2=-\frac14$. There are no 
tachyons in the $\theta^2$ sector since $2v$ is supersymmetric. However, 
the states in this sector do not fill supersymmetric multiplets.
In type IIB, the tachyonic and massless matter includes
\beq
\begin{array}{ll}
\theta^0 : & 14(0,0) + 2(0,1) + 4(1,0) + 8(0,\oh) \\[0.2ex]
\theta, \, \theta^3: & 8(0,0)^- + 40(0,0) + 8(1,0) + 32(\oh,0) \\[0.2ex]
\theta^2: & 50(0,0) + 10(0,1) + 24(0,\oh)
\end{array}
\label{spz4b}
\eeq
Notice that this massless spectrum is anomaly-free since the number of 
$(0,1)$ and $(1,0)$ tensors is the same and the number of $(0,\oh)$ and 
$(\oh,0)$ fermions is the same.
We recall that in the supersymmetric $T^4/\Z_4$, the massless spectrum
consists of a gravity multiplet and 21 tensor multiplets, exactly as in 
type IIB compactification on a smooth K3. In the case at hand, 
the massless states in the $\theta$ sector seem to fill tensor multiplets but 
the states differ in the twisted oscillators acting on them.
In type IIA, the tachyonic and massless matter includes instead
\beq
\begin{array}{ll}
\theta^0 : & 8(0,0) + 6(\oh,\oh) + 4(0,\oh) + 4(\oh,0) \\[0.2ex]
\theta, \, \theta^3: & 8(0,0)^- + 32(0,0) + 8(\oh,\oh) + 16(\oh,0) + 
16(0,\oh) \\[0.2ex]
\theta^2: & 40(0,0) + 10(\oh,\oh) + 12(0,\oh) + 12(\oh,0)
\end{array}
\label{spz4a}
\eeq
In the supersymmetric orbifold the massless spectrum has one 
gravity multiplet and 20 vector multiplets as in a K3 compactification.

Acting on the internal coordinates, the non-supersymmetric $\Z_4$ has
the same action as the supersymmetric one with $v=(0,0,\frac14,-\frac14)$.
We may thus suspect that the non-supersymmetric orbifold is related
to a non-supersymmetric compactification on K3. The same observation
applies to the $T^4/\Z_6$ with $v=(0,0,\frac16,\frac56)$. 

Yet another example that can be related to a K3 compactification
is the $T^4/\Z_6$ with $v=(0,0,\frac13,\frac23)$. In this case there are
no fermions in the spectrum. Besides the untwisted tachyon,
there are 18 twisted tachyons. 
The massless matter in the type IIB (type IIA) orbifold includes
128 (80) singlets and 24 antisymmetric tensors (48 vectors).
Now, this $\Z_6$ is of the form $\Z_3 \times \Z_2$ where the $\Z_3$
has the supersymmetric twist $v=(0,0,\frac13,-\frac13)$ and the $\Z_2$
has the $(-1)^{F_S}$ twist $v=(0,0,0,1)$. Hence, we expect this orbifold
to correspond to a K3 compactification of type 0 strings. Indeed, counting
the zero modes in K3 of the massless type 0 fields we find the same
massless matter as in the orbifold. However, it is not clear how
the 18 twisted tachyons would appear in type 0 on
K3 or disappear in the orbifold upon blowing up the singularities. 
In supersymmetric type II $T^4/\Z_N$ orbifolds
the massless spectrum coincides directly with the K3 spectrum.

Finally let us discuss the $T^4/\Z_5$ that has no supersymmetric analog.
In type IIB the tachyonic and massless matter is given by
\beq
\begin{array}{ll}
\theta^0 : & 8(0,0) + 2(0,1) + 2(1,0) + 4(0,\oh) + 4(\oh,0) \\[0.2ex]
\theta, \, \theta^4: & 10(0,0)^- + 20(0,0) + 10(1,0) + 20(\oh,0) \\[0.2ex]
\theta^2, \, \theta^3: & 10(0,0)^- + 20(0,0) + 10(0,1) + 20(0,\oh)
\end{array}
\label{spz5b}
\eeq
In type IIA we instead find
\beq
\begin{array}{ll}
\theta^0 : & 4(0,0) + 4(\oh,0) + 4(0,\oh) + 4(\oh,\oh) \\[0.2ex]
\theta, \, \theta^4: & 10(0,0)^- + 10(0,0) + 10(\oh,\oh) + 10(\oh,0) + 
10(0,\oh) \\[0.2ex]
\theta^2, \, \theta^3: & 10(0,0)^- + 10(0,0) + 10(\oh,\oh) + 10(\oh,0) + 10(0,\oh)
\end{array}
\label{spz5a}
\eeq

\subsection{$\boldmath{d=4}$}

Now the massless states are classified by the little group $SO_2$, 
{\it i.e.} by helicity $\lambda$. For a $L\otimes R$ state, 
$\lambda=\lambda_r - \lambda_p$ where $\lambda_r$ 
can be read from the first component of the $SO_8$ weight
$r$, and likewise for $\lambda_p$. 
In the untwisted sector, as usual, we find the metric, dilaton and 
antisymmetric tensor and now also a graviphoton. The remaining matter depends
on the $\Z_N$.

As an example, we take the $\Z_6$ with $v=(0,\frac16,\frac16,\frac23)$.
In the sectors $\theta^5$ and $\theta$ there appear tachyons
with $m_R^2= m_L^2=-\frac16$ and multiplicity three (one per
each fixed point). There are no tachyons in other twisted
sectors since $2v$ and $3v$ are supersymmetric. However, the matter 
in these sectors do not fill supersymmetric multiplets. 
Let us now consider type IIB more specifically. In the $\theta^3$
sector there further appear 5 massless vectors. We recall
that in the supersymmetric case with $v=(0,\frac16,\frac16,-\frac13)$
there are 5 vector multiplets in this sector. The remaining
massless states, labelled by helicity, are 
\beq
\begin{array}{ll}
\theta^0 : & 11(0) + 8(\oh)  \\[0.2ex]
\theta^5: & 15(0) + 12(\oh) \\[0.2ex]
\theta^2: & 30(0) + 24(\oh) \\[0.2ex]
\theta^3: & 18(0) + 20(\oh) 
\end{array}
\label{spz6b}
\eeq
In all the \deq4 spectra we do not include the antiparticles. 
In type IIA we find extra massless vectors, five in the untwisted
sector, 3 in the $\theta + \theta^5$, 15 in $\theta^2 + \theta^4$
and 6 in $\theta^3$. The remaining massless states are
\beq
\begin{array}{ll}
\theta^0 : & 6(0) + 4(-\oh) + 4(\oh)  \\[0.2ex]
\theta^5: & 12(0) + 6(-\oh) + 6(\oh) \\[0.2ex]
\theta^2: & 15(0) + 12(-\oh) + 12(\oh) \\[0.2ex]
\theta^3: & 17(0) + 10(-\oh) + 10(\oh) 
\end{array}
\label{spz6a}
\eeq
This $T^6/\Z_6$ orbifold can be seen as a singular Calabi-Yau
with $h_{11}=29$ and $h_{12}=5$. Besides the gravity multiplet
and the dilaton hypermultiplet, the type IIA supersymmetric
compactification has then 29 vector multiplets and 5 hypermultiplets.

We now discuss the $T^6/\Z_6$ with $v=(0,\frac13,\frac13,\frac13)$
so that $\Z_6$ is of the form $\Z_3 \times \Z_2$ where the $\Z_3$
has the supersymmetric twist $v=(0,\frac13,\frac13,-\frac23)$ and 
the $\Z_2$ has the $(-1)^{F_S}$ twist $v=(0,0,0,1)$. Hence, we expect this 
orbifold to be related to type 0 compactification on the singular $T^6/\Z_3$
Calabi-Yau with $h_{11}=36$ and $h_{12}=0$. In this orbifold there
are no twisted tachyons, only the untwisted tachyon that appears
in the $\theta^3$ sector. Besides the graviton, the massless spectrum
in the type IIB (IIA) orbifold includes 222 (78) singlets and 2 (74) vectors.
On the other hand, counting zero modes on a CY of the type 0B massless
fields gives $(6 + 6h_{11} + 2h_{12})$ singlets and $(2 + 2h_{12})$
vectors. For type 0A there are instead $(6 + 6h_{12} + 2h_{11})$
singlets and $(2 + 2h_{11})$ vectors, as one might guess from mirror symmetry.
Thus, the full spectrum of this non-supersymmetric $T^6/\Z_6$ coincides
with the type 0 compactification.

\section{Heterotic}

To study heterotic strings we need to describe the embedding
in the gauge degrees of freedom. We use the bosonic formulation
and realize the embedding by a shift vector $V$ such that
$NV \in \Gamma$, where $\Gamma$ is either the $E_8 \times E_8$
or the $Spin(32)/\Z_2$ lattice. Modular invariance of the partition 
function imposes $N \sum_I V_I = 0\ {\rm mod}\  2$ but this is trivially
satisfied. Modular invariance further requires
\beq
N(V^2 - v^2) = 0\ {\rm mod}\  2
\label{famod}
\eeq
For instance, in $\Z_2$ we can take $V=0$ that leads back to the 
original heterotic string \cite{dh}. Indeed, in some cases supersymmetry 
can be recovered because missing partners appear in the extra twisted 
sectors. More generally, tachyons can be eliminated from the 
spectrum by level matching. The prime example is the non-supersymmetric 
non-tachyonic $SO_{16} \times SO_{16}$ heterotic string in ten dimensions.

In the $E_8 \times E_8$ case, to determine the corresponding shift 
vectors $V$, and the corresponding unbroken gauge group, we use the 
method of deleting nodes in the extended Dynkin diagram \cite{nodes}. 
For $Spin(32)/\Z_2$ we use a similar procedure. All $\Z_2$ embeddings
were given in ref.~\cite{dh} and are conveniently collected in 
Table~\ref{tabz2} (the notation, say $1^{n}$, means that the 
entry 1 is repeated $n$ times). The number of allowed $V$'s grows rapidly 
with $N$ so that we will refrain from giving complete lists. In the following 
we will instead describe some selected examples briefly.

\begin{table}[htb!]
\small
\renewcommand{\arraystretch}{1.25}
\begin{center}
\begin{tabular}{c|c|c|}
\cline{2-3}
{} & $2V$ & Gauge Group \\
\cline{2-3}
{} & $(2,0^7)\times (0^8)$ & $SO_{16} \times E_8$ \\
$E_8 \times E_8$ & $(1^2,0^6)\times (1^2,0^6)$ &
$E_7 \times SU_2 \times E_7 \times SU_2$ \\
{} & $(2,0^7)\times (2,0^7)$ & $SO_{16}\times O_{16}$ \\
\cline{2-3}
{} & $(1^8,0^8)$ & $SO_{16} \times SO_{16}$ \\
$Spin(32)/\Z_2$ & $(1^4,0^{12})$ & $SO_{24} \times SO_8$ \\
{} & $(2,0^{15})$ & $SO_{32}$ \\
{} & $((\oh)^{16})$ & $U_{16}$ \\
\cline{2-3}
\end{tabular}
\end{center}
\caption{$\Z_2$ embeddings and groups.}
\label{tabz2}
\end{table}

We have searched systematically for tachyon-free models for
the $\Z_N$, with $N \leq 5$. By analyzing the mass formula for left
movers (\ref{unolh}) we find that for some, typically larger $V^2$, 
the putative tachyons cannot have $m_R^2=m_L^2$ and hence 
disappear by level matching. Tachyons can also be eliminated by
the generalized orbifold projection described in the appendix. 
Concrete examples are described below.

In the following we will also provide some tachyonic examples.
In particular we will consider the standard embedding 
$V=(v_1,v_2,v_3, 0,\cdots, 0)$ that always leads to models with tachyons.

\subsection{$\boldmath{d=8}$}

We consider first the $T^2/\Z_3$ with $v^2=\frac49$.
In Table~\ref{tabz3} we display the five $E_8 \times E_8$ 
allowed shifts. There
are six embeddings in $Spin(32)/\Z_2$ that lead to gauge group of the
form $U_{3k+1} \times SO_{30-6k}$ and that have $3V$ of type
$(1^{3k+1}, 0^{15-3k})$ or $(1^{3k},-2,0^{15-3k})$.

\begin{table}[htb!]
\small
\renewcommand{\arraystretch}{1.25}
\begin{center}
\begin{tabular}{|c|c|}
\hline
$3V$ & Gauge Group \\
\hline
$(2,0^7)\times (0^8)$ &
$SO_{14} \times U_1 \times E_8$ \\
$(1^2,0^6)\times (1^2,0^6)$ &
$E_7 \times U_1 \times E_7 \times U_1$ \\
$(-2,1^2,0^6)\times (2,0^7)$ &
$E_6 \times SU_3 \times SO_{14} \times U_1$ \\
$\oh (-5,1^7)\times (1^2,0^6)$ &
$SU_9 \times E_7  \times U_1$ \\
$\oh (-5,1^7)\times  \oh(-5,1^7)$ &
$SU_9 \times SU_9$ \\
\hline
\end{tabular}
\end{center}
\caption{$\Z_3$ embeddings and groups in $E_8\times E_8$.}
\label{tabz3}
\end{table}

As an example we consider the standard embedding. In $E_8\times E_8$ the 
resulting gauge group is $SO_{14} \times U_1 \times E_8$ and the matter 
spectrum, including tachyons, is
\beq
\begin{array}{ll}
\theta^0 : & (\r1 + \r4) [(\r{14},\s2) + (\r{64}_s, -\frac1{2\sqrt2}) + 
(\r1,0) ]  \\[0.2ex]
\theta: &  3 (\r1^-) [(\r{14},-\frac1{3\sqrt2}) +(\r1,\frac2{3\sqrt2}) ]
 \\[0.5ex]
{} & +3 (\r1 + \r4) [ (\r{14},-\frac1{3\sqrt2}) + (\r{64}_s, \frac1{6\sqrt2}) 
+ (\r1,-\frac4{3\sqrt2}) + 2(\r1,\frac2{3\sqrt2}) ]
\end{array}
\label{spz3hu}
\eeq
Here $(\r1 + \r4)$ indicates scalar plus fermion and $(\r1^-)$ 
tachyon. Together with the $SO_{14}$ representations we also
give the $U_1$ charge $Q$. 
In $Spin(32)/\Z_2$ the gauge group is $SO_{30} \times U_1$ and
the spectrum is
\beq
\begin{array}{ll}
\theta^0 : & (\r1 + \r4) [(\r{30},\s2) + (\r1,0) ]  \\[0.5ex]
\theta: &  3 (\r1^-) [(\r{30},-\frac1{3\sqrt2}) +(\r1,\frac2{3\sqrt2}) ] \\[0.3ex]
{} & + 3 (\r1 + \r4) [(\r{30},-\frac1{3\sqrt2})  + (\r1,-\frac4{3\sqrt2})
+ 2(\r1,\frac2{3\sqrt2}) ]
\end{array}
\label{spz3hd}
\eeq
In all the \deq8 heterotic spectra such as the above 
antiparticles are not included.
Given the results in the appendix, it is straightforward to
compute the spectrum for all $V$'s. We find that
there is only one tachyon-free model with
gauge group $SU_9 \times SU_9$ and massless matter given by
\beq
\begin{array}{ll}
\theta^0 : & (\r1 + \r4) [(\r1,\r1) + (\r{84},\r1) + (\r1,\r{84}) ]  \\[0.2ex]
\theta: &  3 (\r1 + \r4) (\r9,\r9)
\end{array}
\label{spz3ht}
\eeq

For the $T^2/\Z_4$ with $v^2=\frac14$ there are 14 embeddings in 
$E_8 \times E_8$ and 17 in $Spin(32)/\Z_2$. There are only two tachyon-free 
models in which potential tachyons are eliminated by the generalized 
orbifold projection. One model, with 
$V=\frac18(-7,1^7) \times \frac18(-7,1^7)$, 
has group $(SU_8 \times SU_2)^2$  and massless matter given by
\beq
\begin{array}{ll}
\theta^0 : & (\r1) [(\r{70},\r1,\r1,\r1) + (\r1,\r1,\r{70},\r1) + 
2(\r1,\r1,\r1,\r1)] + \\[0.3ex]
{} & (\r4) [(\r{28},\r2,\r1,\r1) + (\r1,\r1,\r{28},\r2) ]  \\[0.5ex]
\theta: &  4 (\r4)  (\r8,\r1,\r8,\r1)  \\[0.2ex]
\theta^2: &  (\r4) [(\r{28},\r1,\r1,\r2) + (\r1,\r2,\r{28},\r1) ] 
\end{array}
\label{spz2htu}
\eeq
The other non-tachyonic model has $V=\frac14(1^{12},0^4)$, group 
$SU_{12}\times SO_8 \times U_1$ and the following massless matter 
\beq
\begin{array}{ll}
\theta^0 : & (\r1)  [(\r{66},\r1,\frac1{\sqrt6}) + 
(\overline{\r{66}},\r1,-\frac1{\sqrt6})
+ 2(\r1,\r1,0)] + (\r4) (\r{12},\r{8}_v,\frac1{2\sqrt6})   \\[0.2ex]
\theta: &  4 (\r4)  [(\r{66},\r1,-\frac1{2\sqrt6})
+ (\r1,\r8_s,\frac3{2\sqrt6}) + 2(\r1,\r1,-\frac3{2\sqrt6})]  \\[0.2ex]
\theta^2: &  (\r4) (\r{12},\r8_c,\frac1{2\sqrt6})
\end{array}
\label{spz2htd}
\eeq
Notice that in the above $\Tr Q=0$.

We have verified anomaly factorization in all models. 
To compute the anomaly polynomial we just need the contribution of
a $\r4$ fermion in a representation $R_i$ of the gauge group. This is
given by
\beq
{\cal A}_4 = \frac15 \Tr_i F^5 - \frac1{12} \tr R^2 \Tr_i F^3
+ \frac1{240} \tr R^4 \Tr_i F + \frac1{192} (\tr R^2)^2 \Tr_i F 
\label{anomo}
\eeq
We see then that factorization requires cancellation of the
irreducible $\tr F^5$ for all non-Abelian factors. Moreover it
must be that $\Tr Q=0$ and $12\Tr Q^5 = 5\Tr Q^3$ for all $U_1$ factors.
Summing the contribution of all massless fermions
in each model we find that the anomaly polynomial factorizes as
\beq
{\cal A} = \frac1{12} \Tr_{all} {\bf F}^3 \
(\sum_a v_a \tr F_a^2 - \tr R^2)
\label{a8d}
\eeq
Here $a$ runs over all gauge factors and $\tr F_a^n$ stands for trace in 
the fundamental representation, whereas $\Tr_{all} {\bf F}^3$ stands
for trace of all gauge factors over all massless fermions. The
coefficients $v_a$ depend only on the group. Using the conventions of
ref.~\cite{erler} they are $v_a=2,1,1,\frac13,\frac16,\frac1{30}$ for
$SU_N$, $U_1$, $SO_N$, $E_6$, $E_7$ and $E_8$ respectively. To prove
eq.~(\ref{a8d}) we use the results of ref.~\cite{erler}
together with identities such as
\beqa
\Tr_{a_{ij}} F^5 & = & (N-16) \tr F^5 + 10 \tr F^2 \tr F^3 
\nonumber \\[0.2ex]
\Tr_{a_{ijk}} F^5 & = & \frac12 (N-6)(N-27) \tr F^5 + 10 (N-6) \tr F^2 \tr F^3 
\label{trfc}
\eeqa
for antisymmetric $SU_N$ representations with $N \geq 5$.

\subsection{$\boldmath{d=6}$}

We study first the $T^4/\Z_4$.
The generator with $v=(0,0,\frac14,\frac34)$ has the same $V$'s 
that appear in the supersymmetric $v=(0,0,\frac14,-\frac14)$. There
are 12 in $E_8\times E_8$ and 14 in $Spin(32)/\Z_2$.
In the latter case the resulting gauge groups can be of the 
form $U_m \times SO_{2n} \times SO_{32-2m-2n}$, with 
$m+4n-2=0\, {\rm mod} \, 8$. These arise from $4V$ with
vector structure of type $(1^m, 2^n, 0^{16-m-n})$. It is also
possible to obtain gauge group $U_{15-2k} \times U_{2k+1}$ that
result from $4V$ without vector structure of type
$\oh(1^{15-2k}, (-3)^{2k+1})$.

For the standard embedding in $E_8 \times E_8$
the resulting gauge group is $SO_{12} \times SU_2 \times U_1 \times E_8$
and the charged $m^2 \leq 0$ matter is
\beq
\begin{array}{ll}
\theta^0 : & [2(0,0) + (0,\oh)]
[(\r{12},\r2,-\oh) + (\r{32}_c,\r1,\oh) + 2(\r1,\r1,0) + c.c.] + 
\\[0.3ex]
{} & 2(\oh,0)[(\r{32}_s,\r2,0) + (\r1,\r1,1) + (\r1,\r1-1)] \\[0.5ex]
\theta, \, \theta^3: & 4(0,0)^- [(\r{12},\r1,-\frac14) + 
2(\r1,\r2,\frac14) + c.c.] + \\[0.3ex]
{} & 4 [2(0,0) + (\oh,0)] [(\r{32}_s,\r1,\frac14) + (\r1,\r2,-\frac34) + 
2(\r{12},\r1,-\frac14) + 3(\r1,\r2,\frac14) + c.c.]\\[0.5ex]
\theta^2: & (0,\oh)[10(\r{32}_c,\r1,0) + 6(\r{12},\r2,0) + 
32(\r1,\r1,\oh) + 32(\r1,\r1,-\oh)] +\\[0.2ex]
{} & 2(0,0) [10(\r{12},\r2,0) + 6(\r{32}_c,\r1,0) + 
32(\r1,\r1,\oh) + 32(\r1,\r1,-\oh)]
\end{array}
\label{spz4h1}
\eeq
In the $SO_{32}$ heterotic the standard embedding has gauge group
$SO_{28} \times SU_2 \times U_1$ and the following tachyonic 
plus massless matter
\beq
\begin{array}{ll}
\theta^0 : & [2(0,0) + (0,\oh)][(\r{28},\r2,\oh) +  2(\r1,\r1,0) + c.c.] + \\[0.3ex]
{} & 2(\oh,0)[(\r1,\r1,1) + (\r1,\r1-1)] \\[0.5ex]
\theta: & 4(0,0)^- [(\r{28},\r1,\frac14) + 2(\r1,\r2,-\frac14) + c.c.] + \\[0.3ex]
{} & 4 [2(0,0) + (\oh,0)] [2(\r{28},\r1,\frac14) + 3(\r1,\r2,-\frac14) + 
(\r1,\r2,\frac34) + c.c.]\\[0.5ex]
\theta^2: & (0,\oh)[6(\r{28},\r2,0) + 
32(\r1,\r1,\oh) + 32(\r1,\r1,-\oh)] +\\[0.3ex]
{} & 2(0,0) [10(\r{28},\r2,0) + 32(\r1,\r1,\oh) + 32(\r1,\r1,-\oh)]
\end{array}
\label{spz4hd}
\eeq
In this $\Z_4$ we find several non-tachyonic examples. In both heterotic
strings there are models with group $SO_{12} \times U_2 \times SO_{16}$.
There is also a model with group $SO_{10} \times SO_{10} \times U_6$
and another with $SO_{10} \times SU_4 \times SU_8 \times SU_2$ and
massless matter
\beq
\begin{array}{ll}
\theta^0 : & [2(0,0) + (0,\oh)][(\r{16},\r4,\r1,\r1) + (\r1,\r1,\r{28},\r2) +
2(\r1,\r1,\r1,\r1) + c.c.] 
+ \\[0.3ex]
{} & 2(\oh,0)[(\r{10},\r6,\r1,\r1) + (\r1,\r1,\r{70},\r1)] \\[0.5ex]
\theta, \theta^3: & 4 [2(0,0) + (\oh,0)] (\r1,\r4,\r8,\r1) + c.c.] \\[0.3ex]
\theta^2: & (0,\oh)[10(\r{10},\r1,\r1,\r2) + 6(\r1,\r6,\r1,\r2)] 
+ 2(0,0) [6(\r{10},\r1,\r1,\r2) + 10(\r1,\r6,\r1,\r2)]
\end{array}
\label{spz4ht}
\eeq
This model has $V=\frac14(-3,1^3,0^4) \times \frac18(-7,1^7)$.

We next discuss the $T^4/\Z_6$ with $v=(0,0,\frac13,\frac23)$.
There is a large number of allowed embeddings, a few of which with 
$3V^2=\frac23\, {\rm mod}\, 2$ and $3V \in \Gamma$ give back supersymmetric
models. Although we have not searched systematically for 
tachyon-free models we have found a particular non-tachyonic example 
with group $SO_{12} \times SU_2 \times SO_{16} \times U_1$ that can be
obtained in both heterotic strings. For instance, in the $SO_{32}$
heterotic with  embedding $V=\frac16(3^6,1^2,0^8)$ the resulting spectrum is
\beq
\begin{array}{ll}
\theta^0 : & 2(0,0) [(\r{12},\r2,\r1,-\oh) + (\r1,\r1,\r1,1) + 
2(\r1,\r1,\r1,0) + c.c.] + \\[0.3ex]
{} &  (0,\oh)[(\r1,\r2,\r{16},\oh) + c.c.] + 
2(\oh,0)(\r{12},\r1,\r{16},0) \\[0.5ex]
\theta, \, \theta^5: & 9(\oh,0) [(\r{32}_s,\r1,\r1,\frac16) + c.c.] \\[0.5ex]
\theta^2, \, \theta^4: & 
18(0,0) [(\r{12},\r2,\r1,-\frac16) + 2(\r1,\r1,\r1,-\frac23) + 
5(\r1,\r1,\r1,\frac13) + c.c.] + \\[0.3ex] 
{} & 9(0,\oh)[(\r1,\r2,\r{16},-\frac16) + c.c.] \\[0.5ex]
\theta^3: &  2(0,\oh) [(\r{32}_c,\r2,\r1,0) + (\r1,\r1,\r{128},0)] + \\[0.3ex]
{} & (\oh,0) [(\r{32}_s,\r1,\r1,-\oh) + (\r{32}_s,\r1,\r1,\oh)]
\end{array}
\label{spz3hnt}
\eeq
As in \deq10, the $E_8 \times E_8$ spectrum follows from
(\ref{spz3hnt}) upon exchanging the $\theta^k$ and $\theta^{\frac{N}{2}-k}$
fermions. 
As explained in section 3.2, in this case $\Z_6 = \Z_3 \times \Z_2$ 
where the $\Z_3$ respects supersymmetry and $\Z_2$ is equivalent to
$(-1)^{F_S}$. Thus, this orbifold is arguably related to K3
compactification of the 10-dimensional non-supersymmetric
heterotic theories. In fact, we claim that the model (\ref{spz3hnt})
can also be obtained from compactification of the tachyon-free 
$SO_{16} \times SO_{16}$ theory on K3. 

To support our claim we first observe that
\beq
SO_{16} \supset SO_{12} \times SU_2 \times SU_2
\label{sobr}
\eeq
Thus, one $SO_{16}$ can be broken to $SO_{12} \times SU_2$ by an $SU_2$
background with instanton number $k=24$. This is the `standard
embedding' in the $SO_{16} \times SO_{16}$ theory whose bosonic fields 
are $(g_{\mu\nu}, B_{\mu\nu}, \phi)$ and $A_\mu$ in the adjoint
$(\r{120},\r1) + (\r1,\r{120})$. The fermionic fields are $\r8_s$ spinors 
in the $(\r{16},\r{16})$ representation and 
$\r8_c$ spinors in $(\r{128},\r1) + (\r1,\r{128})$. The number of zero modes of
the massless \deq10 fermions that transform in a representation ${\cal R}$
of $G_{10}=SO_{16}$ can be determined from an index theorem \cite{gsw}.
One first decomposes ${\cal R}$ under $G_{10} \supset G \times H$ as 
\beq
{\cal R} = \sum_i  (R_i, S_i)
\label{rdecom}
\eeq
where $R_i$ and $S_i$ denote representations of the unbroken
group $G$ and the bundle group $H$. In the case at hand,
$G=SO_{12} \times SU_2$ and $H=SU_2$ but many other examples can be 
worked out. The basic formula for the number of massless \deq6 fermions 
transforming in the representation $R_i$ is
\beq
n_i = k T(S_i) - {\rm dim}\, S_i
\label{nfer}
\eeq
where $T(S_i)$ is $\Tr_{S_i} T^2$. For $SU_2$, $T(\r2)=\oh$ and $T(\r3)=2$.
We use conventions such that positive (negative) $n_i$ corresponds to 
$(0,\oh)$ ($(\oh,0)$) spinors. Also, recall that in the supersymmetric
case, (\ref{nfer}) gives actually the number of hypermultiplets
which have $2(0,\oh)$. Hence there is an extra factor of two to
be taken into account.
To count the zero modes of the gauge field presumably we can use
supersymmetry as a calculational tool and suppose that 
there exist gauginos in the adjoint. In the supersymmetric
case, gaugino zero modes transforming as $(0,\oh)$
spinors belong to hypermultiplets in which the scalars arise
from the gauge field. Hence, the number of scalar zero modes 
is also given by (\ref{nfer}) taking into account a multiplicity of four
since hypermultiplets have four scalars. Recall further that zero
modes of the metric and antisymmetric tensor give rise to 80
neutral scalars. We will need the following decompositions
\beqa
\r{16} & = & (\r{12},\r1,\r1) + (\r1,\r2,\r2) \nonumber \\[0.2ex]
\r{128} & = & (\r{32}_c,\r2,\r1) + (\r{32}_s,\r1,\r2) \nonumber \\[0.2ex]
\r{120} & = & (\r{66},\r1,\r1) + (\r1,\r3,\r1) + (\r1,\r1,\r3) + (\r{12},\r2,\r2)
\label{vdeco}
\eeqa
Then, (\ref{nfer}) with $k=24$ implies that from the $(\r{16},\r{16})$
fermions there arise  2 $(\oh,0)$ fermions transforming as $(\r{12},\r1,\r{16})$ 
and 20 $(0,\oh)$ fermions transforming as $(\r1,\r2,\r{16})$. This agrees
with (\ref{spz3hnt}) ignoring the $U_1$ charges, {\it i.e.} assuming
that the $U_1$ is broken. Results for the number of fermions in the
$(\r{32}_c,\r2,\r1)$ and $(\r{32}_s,\r1,\r1)$ 
representations also agree in this way.
To obtain the massless scalars, we also use (\ref{nfer}) as explained
above. Then, from the decomposition of the $\r{120}$ we find 40 scalars
transforming as $(\r{12},\r2,\r1)$, same as in the orbifold, and 180 singlets.
Adding the 80 moduli gives 260 singlets as compared to 264 in the
orbifold. Presumably four scalars, corresponding to one hypermultiplet, 
disappear when the $U_1$ is Higgsed away as in the supersymmetric 
compactification.

We might as well consider compactification of the \deq10 tachyonic
models on K3. For example, we can start with the
$SO_{16} \times E_8$ theory whose matter consists of tachyons
transforming as $(\r{16},\r1)$ and massless fermions
$\r8_s$ transforming as $(\r{128}, \r1)$ plus 
$\r8_c$ transforming as $(\overline{\r{128}}, \r1)$ \cite{dh}.
The standard embedding is obtained by embedding an $SU_2$
instanton bundle with $k=24$ in $SO_{16}$ so that the unbroken group is
$SO_{12}\times SU_2 \times E_8$. Using eqs.~(\ref{vdeco}) 
and (\ref{nfer}) one can easily compute the resulting massless content.
Now, we can compare the results with those of the standard
embedding in the $T^4/\Z_6$ with $\Z_6 =\Z_3 \times \Z_2$. In this
orbifold the gauge group is $SO_{12}\times SU_2 \times U_1 \times E_8$
and the massless states match those in the K3 compactificaction
neglecting $U_1$ charges. As expected, in the orbifold there is
an untwisted tachyon transforming as $(\r{12},\r1,0)$ but there
appear furthermore the following twisted tachyons
\beq
\begin{array}{c}
9\, [(\r{12}, \r1,\frac13) + 2(\r1, \r2,-\frac16) + c.c.]
\end{array}
\label{taczt}
\eeq
The states transforming as $(\r1, \r2,-\frac16)$ are of blowing-up type, 
{\it i.e.} they have left-moving oscillators acting on the twisted vacuum. 

We finally consider the $T^4/\Z_5$ orbifold that has no supersymmetric 
analog. Given $v^2=\frac25$
we find 18 distinct embeddings in $E_8 \times E_8$.
In the $SO_{32}$ heterotic there are instead 
20 possibilities that lead to gauge groups of the form
$SO_{32-10n} \times U_{5n}$
and $SO_{30-2j-4n} \times U_{n+j} \times U_{n+1}$, $j=1,6,11$.
There are groups that appear in both heterotic strings, namely
$SO_{12}\times U_5 \times U_5$, $SO_{14}\times U_7 \times U_2$,
$U_8 \times U_8$ and $SO_{10}\times U_3 \times U_8$. In this 
orbifold the shift $V=0$ is allowed. Hence, there are non-supersymmetric
models in six dimensions with group $E_8 \times E_8$ and $SO_{32}$
whose tachyonic and massless spectrum is
\beq
\begin{array}{ll}
\theta^0 : & 4(0,0) + 2(0,\oh) + 2(\oh,0) \\[0.3ex]
\theta, \, \theta^4: & 30(0,0)^- + 50(0,0) + 50(\oh,0) \\[0.3ex]
\theta^2, \, \theta^3: & 30(0,0)^- + 50(0,0) + 50(0,\oh)
\end{array}
\label{spz5hu}
\eeq
All states are neutral. 
In this $\Z_5$ orbifold there is only one tachyon-free 
model with $V=\frac15(-4,1^4,0^3) \times \frac15(-4,1^4,0^3)$,
group $SU_5^4$ and the following massless spectrum
\beq
\begin{array}{ll}
\theta^0 : & [(0,0) + (0,\oh)][(\r5,\r{10},\r1,\r1) +
(\r1,\r1,\r5,\r{10}) + (\r1,\r1,\r1,\r1) + c.c.] 
+ \\[0.3ex]
{} & [(0,0) + (\oh,0)][(\r{10},\r5,\r1,\r1) + (\r1,\r1,\r{10},\r5) + c.c.] \\[0.5ex]
\theta, \theta^4: & 5 [(0,0) + (\oh,0)] [(\r5,\r1,\r5,\r1) + c.c.] \\[0.3ex]
\theta^2, \, \theta^3: & 5 [(0,0) + (0,\oh)] [(\r1,\r5,\r1,\r5) + c.c.]
\end{array}
\label{spz5hd}
\eeq
As a final example, we give the spectrum
for the model with standard embedding in $E_8 \times E_8$. The
gauge group is $SO_{12} \times U_1^2 \times E_8$ and the tachyonic
plus massless matter turns out to be
\beq
\begin{array}{ll}
\theta^0 : & [(0,0) + (0,\oh)][(\r{12},\oh,-\oh) + (\r{32}_c,0,\oh) +
(\r1,-1,0) + (\r1,0,0) + c.c.] 
+ \\[0.3ex]
{} & [(0,0) + (\oh,0)][(\r{12},-\oh,-\oh) + (\r{32}_s,\oh,0) +
(\r1,0,1) + (\r1,0,0) + c.c] \\[0.5ex]
\theta, \theta^4: & 5(0,0)^-[(\r{12},-\frac1{10}, -\frac3{10}) + 
(\r1,-\frac35,\frac15) + 2(\r1,\frac25,\frac15) + c.c.] + \\[0.3ex]
{} & 5 [(0,0) + (\oh,0)] [(\r{32}_s,-\frac1{10},\frac15) + 
(\r{12},-\frac1{10}, -\frac3{10}) + \\[0.3ex]
{} & (\r1,\frac25,-\frac45) + 2(\r1,-\frac35,\frac15) + 
3(\r1,\frac25,\frac15) + c.c.] \\[0.5ex]
\theta^2, \theta^3: & 5(0,0)^-[(\r{12},\frac3{10}, -\frac1{10}) + 
(\r1,-\frac15,-\frac35) + 2(\r1,-\frac15,\frac25) + c.c.] + \\[0.3ex]
{} & 5 [(0,0) + (0,\oh)] [(\r{32}_c,-\frac15,-\frac1{10}) + 
(\r{12},\frac3{10}, -\frac1{10}) + \\[0.3ex]
{} & (\r1,\frac45,\frac25) + 2(\r1,-\frac15,-\frac35) + 
3(\r1,-\frac15,\frac25) + c.c.] 
\end{array}
\label{spz5hse}
\eeq
All states are neutral under $E_8$.

We have checked anomaly factorization in all models. The starting
point is the contribution to the anomaly polynomial due to a $(\oh,0)$
fermion in a representation $R_i$ of the gauge group. This is
\beq
{\cal A}_{(\oh,0)} = 
\frac{{\rm dim}\, R_i}{240}(\tr R^4 + \frac54 (\tr R^2)^2)
+ \Tr_i F^4 - \frac14 \tr R^2 \Tr_i F^2
\label{anoms}
\eeq
For $(0,\oh)$ fermions there is an overall minus sign. Cancellation
of the irreducible $\tr R^4$ requires equal number of $(0,\oh)$ and 
$(\oh,0)$ fermions which is a first easy check on the spectrum.
Cancellation of the irreducible $\tr F^4$ term can also be simply
checked using the results of ref.~\cite{erler} or ref.~\cite{ksty}.
In fact, we have found that in all models the full anomaly polynomial
factorizes as
\beq
{\cal A} =\sum_a \, z_a \tr F_a^2 \
(\sum_a v_a \tr F_a^2 - \tr R^2)
\label{a6d}
\eeq
The coefficients $z_a$ depend on the spectrum of massless fermions
and can be computed using the results in ref.~\cite{erler} or 
in ref.~\cite{ksty} that employs slightly different conventions
such that $v_a=1$ for all groups.

\subsection{$\boldmath{d=4}$}

In the $T^6/\Z_4$ with $v=(0,\oh,\oh,\oh)$ we have looked systematically for 
tachyon-free models. We find that tachyons are eliminated by the orbifold 
projection in examples with groups $U_6 \times U_{10}$, $U_2 \times U_{14}$
and $SU_8 \times SU_2 \times E_6 \times U_2$. There are two
more non-tachyonic examples in this orbifold that have groups
$U_8 \times SO_6 \times SO_{10}$ and $SU_8 \times U_1^2 \times SO_{14}$ 
and can be constructed in both heterotic strings. 

For other $T^6/\Z_N$ we have not performed a systematic search of 
non-tachyonic models, although we have some examples. For instance,
in the $\Z_6$ with $v=(0,\frac16,\frac16,\frac23)$ there are models with
group $SO_{16} \times SO_{10} \times SU_2 \times U_1^2$ in both heterotic
strings. 

In the $\Z_6$ with $v=(0,\frac13,\frac13,\frac13)$ there
are many allowed embeddings, including a few with $3V^2={\rm even}$ 
and $3V \in \Gamma$ that actually produce supersymmetric models.
As explained in section 3.3, this orbifold can be related to compactifications 
of the \deq10 non-supersymmetric theories on the singular $T^6/\Z_3$
Calabi-Yau with $h_{11}=36$ and $h_{12}=0$. Thus, we expect
to find non-tachyonic models arising from the $SO_{16} \times SO_{16}$
theory. Indeed, in both heterotic strings we obtain tachyon-free examples
with group $SO_{16} \times U_6 \times SO_4$, 
$SO_{14} \times U_1 \times SO_{12} \times U_2$ and 
$SO_{16} \times SO_{10} \times SU_3 \times U_1$. In the latter
the massless spectrum turns out to be
\beq
\begin{array}{ll}
\theta^0 : & 3(0) [(\r1, \r{10}, \r3, \frac1{\sqrt6}) ) +
(\r1, \r1, \r3, -\frac2{\sqrt6}) + 3(\r1, \r1, \r1, 0)] + \\[0.3ex]
{} & (\oh) [ (\r{128}, \r1, \r1, 0) + 3(\r1, \r{16}, \r3, -\frac1{2\sqrt6})
+ (\r1, \overline{\r{16}}, \r 1, -\frac3{2\sqrt6}) + 
(\r1, \r{16}, \r1, \frac3{2\sqrt6})]  \\[0.5ex]
\theta: &  27(\oh) (\r{16}, \r1, \r1, -\frac1{\sqrt6}) \\[0.5ex]
\theta^2: &  27(0) [(\r1, \r{10}, \r1, -\frac1{\sqrt6}) )
+ 3(\r1,\r1,\r3,0) + (\r1, \r1, \r1, \frac2{\sqrt6})] 
+ 27(\oh) (\r1, \r{16}, \r1, -\frac1{2\sqrt6}) \\[0.5ex]
\theta^3: & (\oh) [(\r{16}, \r{10}, \r1, 0) + 
3(\r{16}, \r1, \overline{\r3}, -\frac1{\sqrt6})]
\end{array}
\label{spz6hse}
\eeq
This is obtained in $E_8 \times E_8$ with 
$V=(1,0^7)\times \frac16(-3,1^3,0^4)$. In the $SO(32)$ heterotic with
$V=\frac16(0^8,3^5,1^3)$ we obtain the same states.

The model in eq.~(\ref{spz6hse}) should correspond to the standard
embedding in the $SO_{16} \times SO_{16}$ on the $T^6/\Z_3$
Calabi-Yau orbifold. To support this claim we count the number of zero 
modes of the \deq10 fields as in the supersymmetric case.   
We start by decomposing representations under
\beq
SO_{16} \supset SO_{10} \times SU_3 \times U_1
\label{sobrs}
\eeq
The second $SO_{16}$ is broken by a background in $SU_3$ equal
to the spin connection.
Neglecting $U_1$ charges to simplify and already including the
unbroken $SO_{16}$ we have
\beqa
(\r1,\r{120}) & = & (\r1,\r{45},\r1) + (\r1,\r1,\r8) + (\r1,\r1,\r3) +
(\r1,\r1,\overline{\r3}) + (\r1,\r{10},\r3) + (\r1, \r{10},\overline{\r3})
\nonumber \\[0.2ex]
(\r{16},\r{16}) & = & (\r{16},\r{10},\r1) + (\r{16},\r1,\r3) + 
(\r{16},\r1,\overline{\r3}) \nonumber \\[0.2ex]
(\r1,\r{128}) & = & (\r1,\r{16},\r3) + (\r1, \overline{\r{16}}, \overline{\r3}) +
(\r1,\r{16},\r1) + (\r1, \overline{\r{16}}, \r1)  \nonumber \\[0.2ex]
(\r{128}, \r1) & = & (\r{128}, \r1,\r1) 
\label{vdecos}
\eeqa
Since the $(\r{16},\r{10},\r1)$ and the $(\r{128}, \r1,\r1)$ are inert under 
$SU_3$, there is only one zero mode for each that transforms respectively
as $(\r{16},\r{10})$ and $(\r{128}, \r1)$ under $SO_{16} \times SO_{10}$.
This agrees with the orbifold result. From $(\r{16},\r1,\r3) +$ 
$(\r{16},\r1,\overline{\r3})$ there follow instead $h_{11} + h_{12}$
zero modes transforming as $(\r{16},\r1)$. Now, $h_{11}=36$ and $h_{12}=0$
so that the total number coincides with the orbifold spectrum in
eq.~(\ref{spz6hse}) if we count $SU_3$ dimensionality as multiplicity,
{\it i.e.} assuming that $SU_3$ is broken. In fact, the 
$81$ scalars transforming as $(\r1,\r1,\r3,0)$ in the
$\theta^2$ sector are precisely analogous to the blowing-up modes
in the supersymmetric orbifold. Next, recalling that in \deq10 the
$(\r1,\r{128})$ has opposite chirality, we find $h_{11} + 1$ zero
modes transforming as $(\r1,\r{16})$, and $h_{12} + 1$ transforming as 
$(\r1,\overline{\r{16}})$. Notice that the number of families is
$|h_{12} - h_{11}|$. Again, when $SU_3$ is broken, this counting
agrees with the orbifold result. Finally we can also deduce the number
of charged scalars arising from the zero modes of the gauge field. From
the adjoint decomposition we obtain $h_{11} + h_{12}$ states in the
$(\r1,\r{10})$ representation. In the orbifold there are indeed 36
such scalars.

We can as well consider CY compactification of the
tachyonic $SO_{16} \times E_8$ theory. The standard embedding 
leads to group $SO_{10} \times U_1 \times E_8$.
Using (\ref{vdecos}) one can easily compute 
the number of charged zero modes that depend on $h_{11}$ and $h_{12}$.
One can then compare for instance to the standard embedding
in the non-supersymmetric $T^6/\Z_6$ with $v=(0,\frac13,\frac13,\frac13)$
which has group $SO_{10} \times U_1 \times SU_3 \times E_8$.
The charged massless states agree assuming that $SU_3$ is broken
upon repairing the orbifold singularities. Among the orbifold massless
states there are indeed blowing-up scalars. In the orbifold there is
only one untwisted charged tachyon, transforming in the $\r{10}$ of 
$SO_{10}$, that in the CY compactification presumably arises from the 
untwisted \deq10 tachyon transforming as $\r{16}$ of $SO_{16}$.

To verify anomaly factorization we need the 
contribution to the anomaly polynomial due to a Weyl fermion
in a representation $R_i$ of the gauge group. This is
\beq
{\cal A}_{\oh} = \Tr_i F^3 - \frac18 \tr R^2 \Tr_i F
\label{anomc}
\eeq
Cancellation of the irreducible $\tr F^3$ term amounts to
cancellation of non-Abelian cubic anomalies. We find that in 
all models the anomaly properly factorizes as
\beq
{\cal A} = \frac18 \sum_b \, \Tr F_b \
(\sum_a v_a \tr F_a^2 - \tr R^2)
\label{a4d}
\eeq
where $b$ runs only over the $U_1$ factors with $\Tr Q_b \not= 0$.
This factorization requires several non-trivial relations
among the $U_1$ charges such as $\Tr Q_b = 8\Tr Q_b^3$. It can be 
shown that there is at most one combination of $U_1$'s that is anomalous.

\section{Final Comments}

In this work we have used compact orbifolds to construct
tachyonic and non-tachyonic non-supersymmetric models in various 
dimensions. Many more examples can be built. For instance,
one can use $\Z_N \times \Z_M$ orbifolds including discrete torsion
\cite{fiq}. Another interesting extension is to add Wilson lines to 
further break the gauge group \cite{inq}. Connections to non-perturbative 
orbifolds \cite{afiuv} and orbifolds of M-theory \cite{ksty, flo}
can be explored as well.

We have found that some $\Z_N$ non-supersymmetric orbifolds
can be related to K3 and Calabi-Yau (CY) 
compactifications of the type 0 and non-supersymmetric heterotic
strings in \deq10. Indeed, there are tachyon-free orbifolds
that match compactifications of the non-tachyonic 
$SO_{16} \times SO_{16}$ string. Since orbifolds are well defined
string vacua and stable in the absence of tachyons, it then seems that
even without supersymmetry compactification on K3 and CY manifolds
gives consistent vacua at least at tree level. 
The standard embedding in K3 compactification that we discussed appears to 
be one example of a whole class of \deq6 non-tachyonic models whose
massless spectrum can be easily obtained and verified to satisfy
anomaly factorization using basic results. We have also analyzed the
standard embedding in CY and compared in a particular case to the 
corresponding orbifold. 

We might also ask what happens when a \deq10 tachyonic theory is naively
compactified on K3 or a CY and how this relates to a corresponding
orbifold in which the spectrum is exactly known. Concretely, in \deq6
we can compare to the orbifold $T^4/\Z_6$ with $v=(0,0,\frac13,\frac23)$
and in \deq4 to $T^6/\Z_6$ with $v=(0,\frac13,\frac13,\frac13)$.
For type 0 on K3 we found that the massless states agree but there is a 
mismatch in the number of twisted tachyons. For type 0 in CY, results fit 
those of type II on the orbifold in which there is only one untwisted 
tachyon as in type 0. In the heterotic case we studied compactification of
the tachyonic $SO_{16} \times E_8$ theory with standard embedding. 
In \deq6, the orbifold has extra twisted charged tachyons that resemble
blowing-up modes. In \deq4, CY and orbifold results match.

Further orbifold examples associated to compactification of
non-supersymmetric \deq10 theories on K3 and CY can be worked out by 
taking the orbifold point group to be $\Z_N \times \Z_2$ where the 
$\Z_N$ is of supersymmetric type and $\Z_2$ is $(-1)^{F_S}$.
In some orbifolds such as the $T^4/\Z_4$ with $v=(0,0,\frac14,\frac34)$
or the $T^6/\Z_6$ with $v=(0,\frac16,\frac16,\frac23)$, there 
should also be a relation to K3 or CY compactifications in which
supersymmetry is broken. However, the interpretation is not as
simple as in the $\Z_N \times \Z_2$ case.

To conclude we want to stress that there are many other non-supersymmetric 
orbifolds, such as $T^4/\Z_5$, that are not of K3 or CY type. From
the point of view of the orbifold construction all modular invariant
choices of twist and embedding are on the same footing and can be
equally analyzed. In particular, numerous examples can be built and used
to study tachyon condensation. It would be specially interesting
to examine heterotic models.

\vskip1cm
\centerline{\bf Acknowledgements}
\smallskip

A.F. thanks Dieter L\"ust and Ralph Blumenhagen for conversations, 
Stefan Theisen for helpful discussions and observations on the manuscript,
and Luis Ib\'a\~nez and Fernando Quevedo for comments on the manuscript.
A.F. is also grateful to the Institute of Physics of Humboldt 
University, Berlin, for hospitality and the Alexander von Humboldt 
Foundation for support while part of this work was carried out.
This work is partially supported by a CDCH-UCV grant No. 03.214.2001
and a Conicit grant No. G-2001000712.

\vskip1cm


\section{Appendix : Orbifold Spectrum}

To find the spectrum for each model, we follow the
analysis in the supersymmetric case \cite{sierra}.
There are $N$ sectors twisted by $\theta^k, k=0,1,\cdots ,
N-1$.
Each particle state is created by a product of left and right
vertex operators  $L\otimes R$.
At a generic point in the torus moduli
space, the tachyonic and massless right movers follow from
\beq
\label{uno}
m_R^2=N_R+\frac{1}{2}\, (r+k\,v)^2
+E_k-\frac{1}{2} 
\eeq
In this formula $N_R$ is the occupation number of the right-moving 
oscillators and $E_k$ is the twisted oscillator contribution to the
zero point energy given by 
\beq
E_k = \sum_{i=1}^3 \frac{k}2 v_i (1 - kv_i)
\label{evac}
\eeq
When $kv_i > 1$ we must substitute $kv_i \to (kv_i - 1)$ in
(\ref{evac}). The vector $r$ is an $SO_8$ weight.
When $r$ belongs to the scalar or vector class, $r$ takes
the form $(n_0,n_1,n_2,n_3)$, with $n_i$ integer, this is the 
Neveu-Schwarz (NS) sector. When $r$ belongs to an spinorial class
it takes the form $(n_0 + \oh,n_1+ \oh,n_2 + \oh, n_3 + \oh)$,
this is the Ramond (R) sector. The GSO projection turns out
to be $\sum r_a = {\rm odd}$. For example, in the untwisted
sector, the condition $m_R^2=0$ implies $r^2=1$ and the
possible solutions are
\beq 
r=(\underline{\pm 1, 0, 0,0}) = \r8_v
\quad ; \quad
r=\pm (\underline{-\oh, \oh, \oh ,\oh}) = \r8_s 
\label{re1}
\eeq
where underlining means permutations.  
The vector $v$ in (\ref{uno}) is $(0,v_1,v_2,v_3)$, with the $v_i$
given in Table~\ref{tab1}. For simplicity we are setting 
$\alpha^{\prime}=2$ everywhere.

The mass formula for left-movers depends on the type of string. 
For type II we have
\beq
m_L^2  =  N_L + \frac{1}{2}\, (p+k\,v)^2
+ E_k-\frac{1}{2}  
\label{unol}
\eeq
where $p$ is an $SO_8$ weight as well. In type IIB the GSO
projection is also $\sum p_i = {\rm odd}$ in both NS and R sectors.
In type IIA one has instead $\sum p_i = {\rm even}$ in the R sector.
In the untwisted sector the spinor weights are then those of $\r8_c$. 

For the heterotic string in the lattice formulation the left mass 
formula is instead
\beq
m_L^2  = N_L + \frac{1}{2}\, (P+k\,V)^2+ E_k -1 
\label{unolh}
\eeq
where $P$ belongs to the lattice $\Gamma$. The vectors $(P+kV)$
determine the representation of the state under the gauge group.
$N_L$ is the occupation number for all left-moving oscillators.

States in a $\theta^k$-twisted sector with given
$m_R^2=m_L^2$ are characterized by 
$(N_R,r; N_L,p)$ in type II or by $(N_R,r; N_L,P)$ in the
heterotic string. The degeneracy of these states follows from
the generalized orbifold projection given by
\beq
{\cal D}(\theta^k)=\frac{1}{N}\sum_{\ell=0}^{N-1}\chi(\theta^k,\theta^\ell)\,
\Delta(k,\ell) ~,
\label{dos}
\eeq
In the heterotic string the phase $\Delta$ is
\beq
\label{tres}
\Delta(k,\ell)=\exp\{2\pi\, i[(r+kv)\cdot \ell v-
(P+kV)\cdot  \ell V + \oh k\ell (V^2-v^2)+\ell(\rho_L + \rho_R)]\} 
\eeq
In type II the phase $\Delta$ is analogous with $p,v$ instead of $P,V$.
The factors $\rho_{R,L}$ appear only in the case of states with
oscillation number due to the internal (complex) coordinates $Y_i$.
The phase $e^{2\pi\, i\rho}$ indicates how the oscillator is
rotated by $\theta$. In the $\theta$ sector, for example, 
the right-moving oscillators are of the
form $\alpha^i_{-v_i} , \alpha^i_{-1-v_i}, \cdots$ with 
$\rho_R=v_i$ and oscillators ${\bar \alpha}^i_{-1+v_i} , 
{\bar \alpha}^i_{-2+v_i}, \cdots$ with $\rho_R=-v_i$. Similarly,
for left-movers there can be
oscillators ${\bar {\tilde \alpha}}^i_{-v_i} , 
{\bar {\tilde \alpha}}^i_{-1-v_i}, \cdots$ with 
$\rho_L=-v_i$ and oscillators ${\tilde \alpha}^i_{-1+v_i} , 
{\tilde \alpha}^i_{-2+v_i}, \cdots$ with $\rho_L=v_i$.

In (\ref{dos}) $\chi(\theta^n,\theta^m)$ is a numerical factor
that counts the fixed point multiplicity. More concretely,
$\chi(1,\theta^\ell)=1$, so that in the untwisted sector
${\cal D}(1)$ projects out precisely the states non-invariant under 
$\theta$. In twisted sectors $\chi(\theta^k,\theta^\ell)$ 
is the number of simultaneous fixed points of $\theta^k$ 
and $\theta^\ell$. 

As an example, let us work out the type IIB string
on $T^2/\Z_3$ with $v=(0,0,0,\frac23)$. 
In the untwisted sector there are no tachyons and
the massless states are as follows.
With $r\cdot v = p\cdot v= 0$ we have
\beq
\begin{array}{cc}
r &  p \\
(\underline{\pm 1,0,0},0) & (\underline{\pm 1,0,0},0) 
\end{array}
\label{ugbd}
\eeq
The first three entries in $r,p$ correspond to non-compact
coordinates and give the Lorentz representations according
to the little group $SO_6$. The NS-NS states in (\ref{ugbd}) 
are thus $\r6\otimes \r6= \r{20} + \r{15} + \r1$. 
These are the metric, the antisymmetric tensor and the dilaton.
With $r\cdot v = p\cdot v= \frac23 \, {\rm mod \ int}$ we have
\beq
\begin{array}{cc}
r &  p \\
(0,0,0,1) & (0,0,0,1) \\[0.2ex]
(\underline{\oh ,-\oh, -\oh},-\oh) & (\underline{\oh ,-\oh, -\oh},-\oh)
\\[0.2ex]
(\oh,\oh,\oh,-\oh) & (\oh,\oh,\oh,-\oh)
\end{array}
\label{umba}
\eeq
The four spinor weights form the $\r4$ of $SO_6$. Then, for instance
when we combine right and left we obtain $\r1 + \r{15}$
in the R-R sector.
With $r\cdot v = p\cdot v= \frac13 \, {\rm mod \ int}$
the solutions are as in (\ref{umba}) with opposite sign.
In the $\theta$-twisted sector the solutions with $m_R^2 \leq 0$ are
\beq
\begin{array}{ccc}
r &  N_R & m_R^2 \\[0.2ex]
(0,0,0,-1) & 0 & -\frac13 \\[0.2ex]
(\underline{\oh ,-\oh, -\oh},-\oh) & 0 & 0 \\[0.2ex]
(\oh,\oh,\oh,-\oh) & 0 & 0 \\[0.2ex]
(0,0,0,-1) & \frac13 & 0 \\[0.2ex]
\end{array}
\label{tz3}
\eeq
The solutions with $m_L^2 \leq 0$ are similar. 
The solutions in the $\theta^2$-twisted sector are as in 
(\ref{tz3}) with opposite sign and will lead therefore to
the antiparticles of the states in the $\theta$ sector.

In type IIA we have to take the $p$'s arising instead from
$\r8_c$. This implies for example, to change the $p$ spinor weights
in (\ref{umba}) to the $\overline{\r4}$ of $SO_6$.

\end{document}